\documentclass[usenatbib]{mnras}
\usepackage{amsmath}
\usepackage{natbib}
\usepackage{epsfig}
\usepackage{graphicx}
\usepackage{amssymb}
\usepackage{url}
\usepackage{xcolor}

\def\gtsima{$\; \buildrel > \over \sim \;$}
\def\ltsima{$\; \buildrel < \over \sim \;$}
\def\gtrsim{\lower.5ex\hbox{\gtsima}}
\def\lesssim{\lower.5ex\hbox{\ltsima}}

\newcommand{\msun}{$\mathrm{M}_{\odot}$}

\colorlet{nicola}{green!10!orange!90!}

\usepackage{amsmath}	
\usepackage{amssymb}	
\usepackage{mathrsfs}

\begin{document}

\title[BBHs in the PI mass gap]{Binary black holes in the pair-instability mass gap}
\author[]{Ugo N. Di Carlo$^{1,2,3}$, Michela Mapelli$^{4,2,3}$, Yann Bouffanais$^{4,2}$, Nicola Giacobbo$^{4,2,3}$, \newauthor Filippo Santoliquido$^{4,2}$ Alessandro Bressan$^{5}$, Mario Spera$^{2,4,6,7}$ and Francesco Haardt$^{1}$\parbox{\linewidth}{}
\vspace{0.3cm}
\\
$^{1}$Dipartimento di Scienza e Alta Tecnologia, University of Insubria, Via Valleggio 11, I--22100, Como, Italy
\\
$^{2}$INFN, Sezione di Padova, Via Marzolo 8, I--35131, Padova, Italy
\\
$^{3}$INAF-Osservatorio Astronomico di Padova, Vicolo dell'Osservatorio 5, I--35122, Padova, Italy
\\
$^{4}$Dipartimento di Fisica e Astronomia `G. Galilei', University of Padova, Vicolo dell'Osservatorio 3, I--35122, Padova, Italy
\\
$^{5}$Scuola Internazionale Superiore di Studi Avanzati (SISSA), Via Bonomea 265, I-34136 Trieste, Italy
\\
$^{6}$Center for Interdisciplinary Exploration and Research in Astrophysics (CIERA), Evanston, IL 60208, USA
\\
$^{7}$Department of Physics \& Astronomy, Northwestern University, Evanston, IL 60208, USA
}
\maketitle \vspace {7cm}
\bibliographystyle{mnras}

\begin{abstract}
Pair instability (PI) and pulsational PI prevent the formation of black holes (BHs) with mass $\gtrsim{}60$ M$_\odot$ from single star evolution. Here, we investigate the possibility that BHs with mass in the PI gap form via stellar mergers and multiple stellar mergers, facilitated by dynamical encounters in young star clusters. We analyze $10^4$ simulations, run with the direct N-body code {\sc nbody6++gpu} coupled with the population synthesis code {\sc mobse}. We find that up to $\sim{}6$~\% of all simulated BHs have mass in the PI gap, depending on progenitor's metallicity. This formation channel is strongly suppressed in metal-rich ($Z=0.02$) star clusters, because of stellar winds. BHs with mass in the PI gap are initially single BHs but can efficiently acquire companions through dynamical exchanges. We find that $\sim{}21$\%, 10\% and 0.5\% of all binary BHs have at least one component in the PI mass gap at metallicity $Z=0.0002$, 0.002 and 0.02, respectively. Based on the evolution of the cosmic star formation rate and metallicity, and under the assumption that all stars form in young star clusters, we predict  that $\sim{}5$~\% of all  binary BH mergers detectable by advanced LIGO and Virgo at their design sensitivity have at least one component in the PI mass gap. 
\end{abstract}


\maketitle

%

\section{Introduction}
The  mass function of stellar black holes (BHs) is highly uncertain, as it crucially depends on complex physical processes affecting the evolution and the final fate of massive stars. For a long time, we had to rely on a scanty set of observational data, mostly dynamical mass measurements of compact objects in X-ray binaries \citep{ozel2010,farr2011}. In the last four  years, gravitational wave (GW) data have completely revolutionised our perspective: ten binary black holes (BBHs) have been observed during the first and the second observing run of the LIGO-Virgo collaboration (LVC, \citealt{abbottGW150914,abbottO1, abbottO2,abbottO2popandrate}) and we expect that several tens of new BBH mergers will be available 
 as a result of the third observing run. GW data will soon provide a Rosetta Stone to decipher the mass function of BBHs.

Thus, it is particularly important to advance our theoretical understanding of BH formation and BH mass function, in order to provide an interpretative key for future GW data. We currently believe that the mass of a BH depends mainly on the final mass of its progenitor star and on the details of the supernova (SN) explosion (e.g. \citealt{heger2003,mapelli2009,mapelli2010,mapelli2013,belczynski2010,fryer2012,spera2015,limongi2018}). Among all types of SN explosion, pair instability SNe (PISNe) and pulsational pair instability SNe (PPISNe) are expected to leave a strong fingerprint on the mass function of BHs. 
If the He core mass is larger than $\sim{}30$~M$_\odot$, soon after carbon burning when the stellar core temperature reaches $\sim{}7\times{}10^8$~K, effective pair production softens the equation of state, leading to a loss of pressure. The stellar core contracts, triggering neon, oxygen and even silicon burning in a catastrophic way, known as pair instability (PI).  Stars developing a helium core mass $64\le{}m_{\rm He}/M_\odot{}\le{}135$ are thought to be completely disrupted by a PISN, leaving no compact object \citep{heger2003}. Stars with a smaller helium core ($32\lesssim{}m_{\rm He}/M_{\odot}\lesssim{}64$) undergo pulsational PI: they go through a series of pulsations, losing mass with an enhanced rate, till their cores leave the mass range for PI \citep{woosley2007}.

The combination of PISNe and PPISNe leads to a mass gap in the BH mass function between $\sim{}60$ M$_\odot$ and $\sim{}120$ M$_\odot$. Both the lower and the upper edge of the mass gap depend on the details of massive star evolution. In particular, the lower edge of the mass gap might span from $\sim{}40$ M$_\odot$  up to $\sim{}65$ M$_\odot$, depending on the details of PI, stellar evolution and core-collapse SNe \citep{belczynski2016pair,woosley2017,woosley2019,spera2017,giacobbo2018,giacobbo2018b,marchant2019,mapelli2019b,stevenson2019,farmer2019,renzo2020}. The upper edge of the gap is even more uncertain. LIGO-Virgo data from the first and second observing run are consistent with a maximum BH mass of $\approx{}45$ M$_\odot$, in agreement with  the existence of a PI mass gap \citep{abbottO2popandrate}.



However, some exotic BH formation channels might populate the PI gap. Hence, the detection of a BH in the mass gap by the LVC would possibly provide a smoking gun for these exotic channels. Primordial BHs (i.e. BHs formed from the collapse of gravitational instabilities in the early Universe, e.g. \citealt{carr1974,carr2016}) might have a mass in the gap. Alternatively, BHs with mass in the gap can form as ``second-generation'' BHs \citep{gerosa2017}, i.e. BHs born from the merger of two smaller BHs.

Finally, \cite{spera2019} and \cite{dicarlo2019} proposed a third possible channel to produce BHs in the mass gap. If a massive star with a well-developed helium core merges with a non-evolved companion (a main sequence or an Hertzsprung-gap star), it might give birth to an evolved star with an over-sized hydrogen envelope. If the helium core remains below $\sim{}32$ M$_\odot$ and the star collapses to a BH before growing a much larger core and before losing a significant fraction of its envelope, the final BH might be in the PI mass gap. 

If a second-generation BH or a BH born from stellar merger form in the field, they remain single objects and we do not expect to observe them in a BBH merger. In contrast, if they form in a dense stellar cluster they might capture a new companion through a dynamical exchange, possibly becoming a BBH \citep{miller2002,dicarlo2019,rodriguez2019,gerosa2019}. Here, we focus on BHs in the PI gap formed from stellar mergers and we estimate their mass range, merger efficiency and detection probability.


\section{Methods}
The simulations discussed in this paper were done using the same code and methodology as described in \cite{dicarlo2019}. In particular, we use the direct summation N-Body code \textsc{nbody6++gpu} \citep{wang2015} coupled with the new population synthesis code \textsc{mobse} \citep{mapelli2017,giacobbo2018,giacobbo2018b}. {\sc mobse} includes up-to-date prescriptions for massive star winds, for core-collapse SN explosions and for PISNe and PPISNe. 


In this work, we have analyzed the simulations of $10^4$ fractal young star clusters (SCs);
  4000 of them are the simulations presented in \cite{dicarlo2019}, while the remaining 6000 are discussed in \cite{dicarlo2020}. The initial conditions of the simulations presented in this paper are summarized in Table~\ref{tab:table1}.
Unlike globular clusters, young SCs are asymmetric, clumpy systems. Thus, we model them with fractal initial conditions \citep{kuepper2011}, to mimic initial clumpiness \citep{goodwin2004}. The level of fractality is decided by the parameter $D$ (where $D=3$ means homogeneous distribution of stars). In this work, we assume $D=1.6,\,{}2.3$. 

The total mass $M_{\rm SC}$ of each SC (ranging from $10^3$ \msun{} to $3\times{}10^4$ \msun{})  is drawn from a distribution $dN/dM_{\rm SC}\propto M_{\rm SC}^{-2}$, as the embedded SC mass function described in \cite{lada2003}. We choose to simulate SCs with mass $M_{\rm SC}<30000$ \msun{} for computational reasons. Thus, the mass distribution of our simulated SCs mimics the mass distribution of SCs in Milky Way-like galaxies. We choose the initial SC half mass radius $r_{\rm h}$ according to the Marks \& Kroupa relation \citep{marks2012} in 7000 simulations, and we adopt a fix value $r_{\rm h}=1.5$ pc for the remaining 3000 simulations. 

The stars in the simulated SCs follow a \cite{kroupa2001} initial mass function, with minimum mass 0.1 \msun{}  and maximum mass 150 \msun{}. We assume an initial binary fraction $f_{\mathrm{bin}}=0.4$. The orbital periods, eccentricities and mass ratios of binaries are drawn from \cite{sana2012}. We simulate each star cluster for 100 Myr in a rigid tidal field corresponding to the Milky Way tidal field at the orbit of the Sun. We refer to \cite{dicarlo2019} for further details on the code and on the initial conditions.

 We consider three different metallicities: $Z=0.0002,$ 0.002 and 0.02 (approximately 1/100, 1/10 and 1 Z$_\odot$). 
We divide our simulations in three sets, corresponding to metallicity $Z=0.0002$ (2000 runs), 0.002 (6000 runs) and 0.02 (2000 runs). The simulations with $Z=0.002$ are the union of the 4000 runs  presented in \cite{dicarlo2019} and 2000 runs discussed in \cite{dicarlo2020}. The simulations with $Z=0.02$ and $Z=0.0002$ are both from \cite{dicarlo2020}.  The main differences between the simulations already presented in \cite{dicarlo2019} and the new runs from \cite{dicarlo2020} are i) the efficiency of common envelope ejection ($\alpha=3$ in \citealt{dicarlo2019} and $\alpha{}=5$ in \citealt{dicarlo2020}), and ii) the model of core-collapse supernova (the rapid and the delayed models from \citealt{fryer2012} are adopted in \citealt{dicarlo2019} and in \citealt{dicarlo2020}, respectively). 
Putting together these different samples is not a completely consistent approach, but is justified by the fact that the population of BHs with mass in the $60-150$ M$_\odot$ range is not strongly affected by these different assumptions. For example, in \cite{dicarlo2020}, we showed that our different assumptions change the percentage of BHs in the gap by a factor of $\sim{}1.1-1.5$ (this is much less than the impact of stellar metallicity we want to probe here). Finally, putting together different SC models is important to filter out stochastic fluctuations, since the formation of BHs in the gap is a rare event and our simulations are computationally expensive.

\begin{table}
\begin{center}
\caption{\label{tab:table1} Initial conditions.} \leavevmode
\begin{tabular}[!h]{cccccc}
\hline
Set & $Z$ & $N_{\rm sim}$ & $r_{\rm h}$  & $D$ & ref. \\ 
\hline
Z0002 & 0.0002 & 1000  & M2012 & 1.6 &  D2020 \\
 & 0.0002 & 1000  & 1.5 pc           & 1.6 & D2020 \\
 \hline
Z002 & 0.002 & 2000   & M2012 & 2.3 & D2019 \\
 & 0.002 & 3000   & M2012 & 1.6 & D2019, D2020 \\
& 0.002 & 1000   & 1.5 pc           & 1.6 & D2020\\
\hline
Z02 & 0.02  & 1000   &  M2012 & 1.6 & D2020\\
 & 0.02  & 1000   & 1.5 pc            & 1.6 & D2020\\
\hline
\end{tabular}
\end{center}
\begin{flushleft}
\footnotesize{ Column~1: Name of the simulation set. Column~2: metallicity $Z$. Column~3: Number of runs performed per each set. Column~4: half-mass radius $r_{\rm h}$. M2012 indicates that half-mass radii have been drawn according to \cite{marks2012}. Column~5: fractal dimension ($D$). Column~6: reference for each simulation set. D2019 and D2020 correspond to \cite{dicarlo2019} and \cite{dicarlo2020}, respectively.}
\end{flushleft}
\end{table}

\section{Results}
\begin{figure}
  \center{
    \epsfig{figure=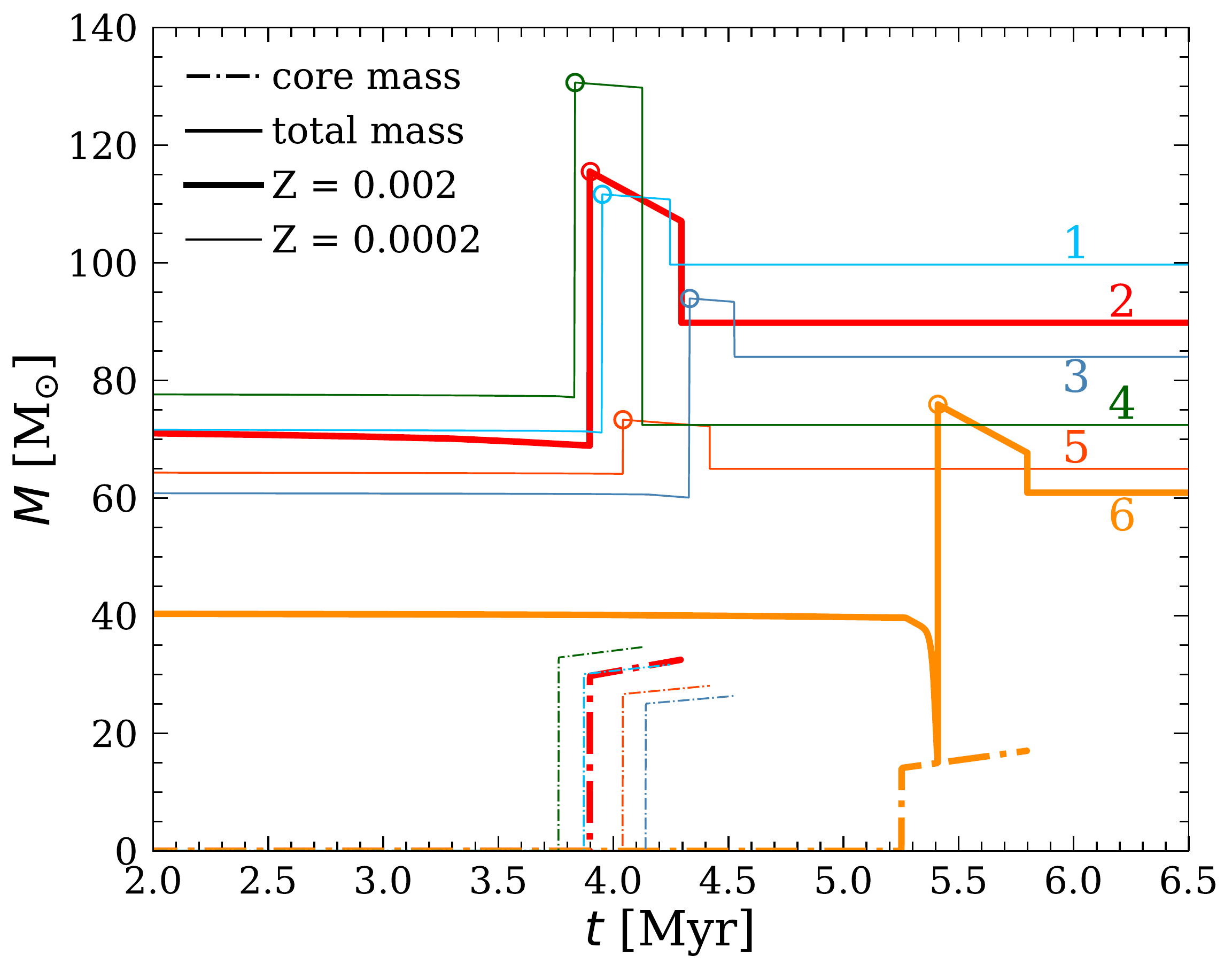,width=8.0cm}

    \caption{\label{fig:evol} Evolution of the total mass (solid lines) and the core mass (dot-dashed lines) of the progenitors of a sample of BHs with mass in the gap. The open circle marks the time of the merger with a companion star. Thick lines: $Z=0.002$; thin lines: $Z=0.0002$. Models 1, 3 and 4 (light blue, blue and green) are stars that become single BHs; models 2, 5 and 6 (red, orange and yellow) are stars which end up in merging BBHs.} 
}
\end{figure}

From our simulations, we extract information on BHs with mass in the PI gap, between 60 and 150 M$_\odot$ (given the uncertainties on the edges of the mass gap, we make a conservative assumption for both the lower and the upper edge of the mass gap). In \cite{dicarlo2019}, we have already discussed the properties of BHs that form from stars with $Z=0.002$, have mass in the PI gap and merge with other BHs in less than a Hubble time. Here, we extend our study to other progenitor's metallicities ($Z=0.02$ and 0.0002), because stellar metallicity is a crucial ingredient to understand how many BHs can form with mass in the PI gap. Moreover, we discuss the formation pathways of BHs born from stellar mergers, by looking at the core and envelope evolution of their progenitors (Figure~\ref{fig:evol}). 
We consider all BHs that form in the PI mass gap (both single and binary BHs) and we investigate their properties. Finally, we estimate the detectability of BHs in the mass gap by LIGO and Virgo at design sensitivity. 


\subsection{Formation channels of BHs in the gap}
The vast majority of BHs with mass in the PI  gap that form in our simulation originates from the merger of an evolved star (with a developed helium core of mass $\approx{}15-30$ M$_\odot$) and a main sequence companion. The merger is generally triggered by dynamical perturbations. In several cases, the evolved star is the result of multiple mergers between other stars, facilitated by the dense dynamical environment. This process of multiple mergers occurring in a very short time span is known as runaway collision and was already discussed in several papers (see e.g. \citealt{portegieszwart2002,portegieszwart2004,giersz2015,mapelli2016,gieles2018}). 

 Figure~\ref{fig:evol} shows the evolution of six stellar progenitors of BHs in the PI mass gap. Three of these BHs become members of BBHs and merge within a Hubble time, while the other three objects leave single BHs. We find no significant difference between the formation channel of merging BHs in the PI mass gap and that of single BHs or  non-merging BBHs 
 with mass in the PI gap. 

  The stars shown in Figure~\ref{fig:evol} undergo a merger with a main-sequence companion in their late evolutionary stages ($\sim{}4-6$ Myr), when they are Hertzsprung gap or core helium burning stars. We assume that there is no mass loss during the merger. 
  The merger products are not significantly rejuvenated, because they already developed a He core. They are evolved based on their mass and are subject to stellar winds, depending on their metallicity. Their final He core is $\sim{}17-32$ M$_\odot$ (below the PPISN/PISN gap), while their hydrogen envelope is over-sized with respect to single star evolution, because of the merger. While most stars in Figure~\ref{fig:evol} simply merge with another star without previous mass transfer episodes, star number~6  shows signature of mass transfer. This star fills its Roche lobe after leaving the main sequence and its hydrogen envelope is removed. At the end of mass transfer, it merges with its companion. 

In all the simulations, the post-merger star evolves for $t_{\rm post-merg}=t_{\rm He}+t_{\rm C}+t_{\rm Ne}+t_{\rm O}+t_{\rm Si}\sim{}t_{\rm He}$, where  $t_{\rm post-merg}$ is the time remaining to collapse, while $t_{\rm He}$, $t_{\rm C}$, $t_{\rm Ne}$, $t_{\rm O}$ and $t_{\rm Si}$ are the timescale of helium, carbon , neon, oxygen and silicon burning, respectively. During $t_{\rm post-merg}$, the star converts a mass $\Delta{}M_{\rm He}\sim{}\dot{M}_{\rm He}\,{}t_{\rm post-merg}$ into heavier elements, where 
\begin{eqnarray}\label{eq:convertCNO}
  \dot{M}_{\rm He}\lesssim{}{}2\times{}10^{-5}\,{}{\rm M}_\odot\,{} {\rm yr}^{-1}\,{}\left(\frac{L_\ast{}}{10^6\,{}{\rm L}_\odot}\right)\nonumber{}\\\times{}\left(\frac{6.3\times{}10^{18}\,{}{\rm erg}\,{}{\rm g}^{-1}}{\eta_{\rm CNO}}\right)\,{}\left(\frac{0.5}{X}\right).
\end{eqnarray}
In equation~\ref{eq:convertCNO}, $L_\ast$ is the stellar luminosity, $X$ is the hydrogen fraction and $\eta_{\rm CNO}$ is the efficiency of mass-to-energy conversion during the CNO cycle (e.g. \citealt{prialnik2000}).

  If the final mass of the helium core $M_{\rm He,\,{}f}=M_{\rm He}+\Delta{}M_{\rm He}<32$ M$_\odot$, where $M_{\rm He}$ is the mass of the helium core before the last stellar merger, then the star with an oversized hydrogen envelope can avoid PI and directly collapses to a BH, possibly with mass $>60$ M$_\odot$. This is just an order of magnitude estimation, more refined calculations would require a hydrodynamical simulation to follow the merger (see e.g. \citealt{gaburov2010}) and a stellar-evolution code to integrate nuclear burning and stellar evolution.



Once they form, BHs with mass in the gap are efficient in acquiring companions:  $\sim{}21$~\% and $\sim{}10$~\% of all BBHs have at least one member with mass in the PISN gap at $Z=0.0002$ and $Z=0.002$, respectively. This is expected, because these BHs are significantly more massive than the other BHs and stars in the SCs, and dynamical exchanges favour the formation of more massive binaries, which are more energetically stable (see e.g. \citealt{hills1980}). 

If we consider only BBHs merging within a Hubble time (14 Gyr) due to GW emission, only $\sim{}2.2$~\% and $\sim{}2.1$~\% of them have at least one BH in the PI gap at $Z=0.0002$ and $Z=0.002$, respectively. We find only 11 merging BBHs with a BH in the PI gap, hence these percentages are affected by stochastic fluctuations (see Table~\ref{tab:table2} for an estimate of the uncertainties). These BBHs merge after being ejected from their parent young SC. Finally, we find no merging BBHs with members in the PI gap at solar metallicity.

None of the BBHs in our simulations hosts a second-generation BH (i.e. a BH that forms from the merger of two BHs). The low escape velocity from our SCs (up to few km s$^{-1}$ in the most massive SCs) prevents second-generation BHs from remaining inside the cluster: all of them are ejected and cannot acquire a new companion. In contrast, in massive SCs (like globular clusters and nuclear star clusters) second-generation BHs have a significantly higher chance of remaining inside their parent cluster and acquiring a companion (see e.g. \citealt{miller2002,colpi2003,antonini2016,rodriguez2019,arcasedda2019,arcasedda2020}).

 It is important to highlight several caveats inherent with our analysis. First, {\sc mobse} assumes that no mass is lost during the merger while hydrodynamical simulations have shown that mass ejecta can represent up to $\sim{}25$\% of the total mass  (\citealt{gaburov2010}, see also \citealt{dale2006,justham2014,vigna2019,wu2020}). We have re-simulated the six objects in Figure~\ref{fig:evol} assuming that all of them lose 25\% of their mass after each merger. The masses of the resulting BHs are lower by $\sim{}22-28$\%; three of the six BHs in Figure~\ref{fig:evol} are still in the mass gap (tracks 1, 2 and 3), while the remaining three have mass $<60$ M$_\odot$. 
 
 Furthermore, the polynomial fitting formulas implemented in {\sc mobse} might be inaccurate to describe the final evolution of such post-merger massive stars. In a follow-up work, we will evolve our post-collision models with a stellar evolution code\footnote{ \cite{glebbeek2009} re-simulated a runaway collision product with a stellar evolution code. They find that mass loss strongly suppresses the formation of massive mergers at solar metallicity, while a final stellar mass $\sim{}260$ M$_\odot$ is possible at $Z=0.001$. This is similar to our findings. However, their results are not directly comparable with ours, because the original $N-$body simulation they start from is composed of 131072 particles; thus, the runaway collision product is significantly more massive than ours.}, to check any deviations from {\sc mobse}.  In addition, we assume that the final hydrogen envelope entirely collapses to a BH. This final outcome depends on the final binding energy of the envelope (see e.g. \citealt{sukhbold2016} for a discussion). Finally, we model PPISNe with a fitting formula \citep{spera2017} to the models by \cite{woosley2017}. However, the models by \cite{woosley2017} are suited for stars following regular single stellar evolution, that could be significantly different from merger products. 

\begin{figure}
  \center{
    \epsfig{figure=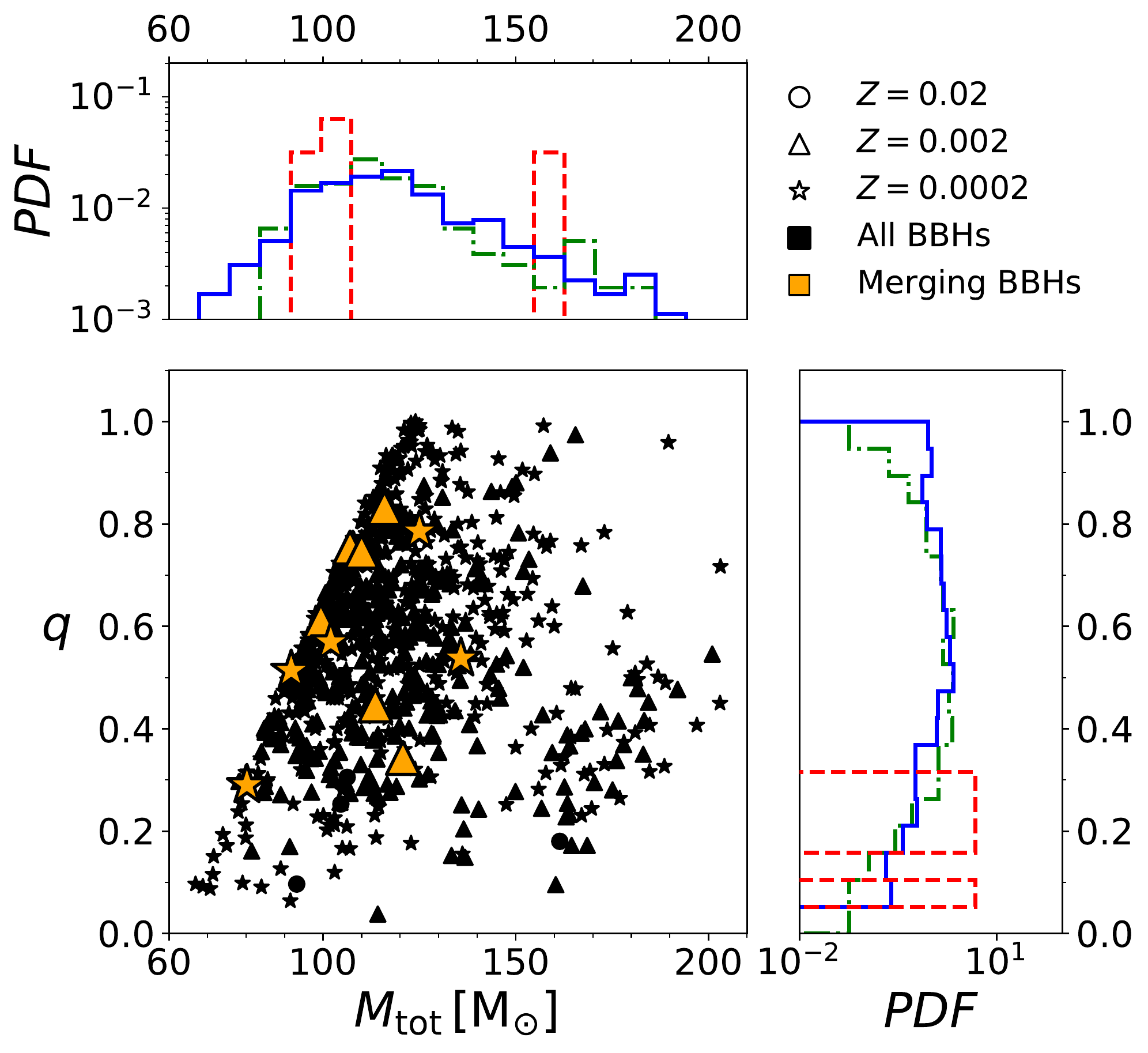,width=8.0cm}
    \caption{\label{fig:qmtot} Mass ratio $q=M_2/M_1$ versus total mass $M_{\rm tot}=M_1+M_2$ of BHs with mass in the gap that are members of BBHs by the end of the simulations. Circles, triangles and stars refer to $Z=0.02$, 0.002 and 0.0002, respectively. Orange and black symbols refer to BBHs merging within a Hubble time and to all BBHs, respectively. Marginal histograms show the distribution of $q$ (on the $y-$axis) and $M_{\rm tot}$ (on the $x-$axis). 
    Solid blue, dot-dashed green and dashed red histograms refer to $Z=0.0002,$ 0.002 and 0.02, respectively.} 
  }
\end{figure}

\begin{figure}
  \center{
    \epsfig{figure=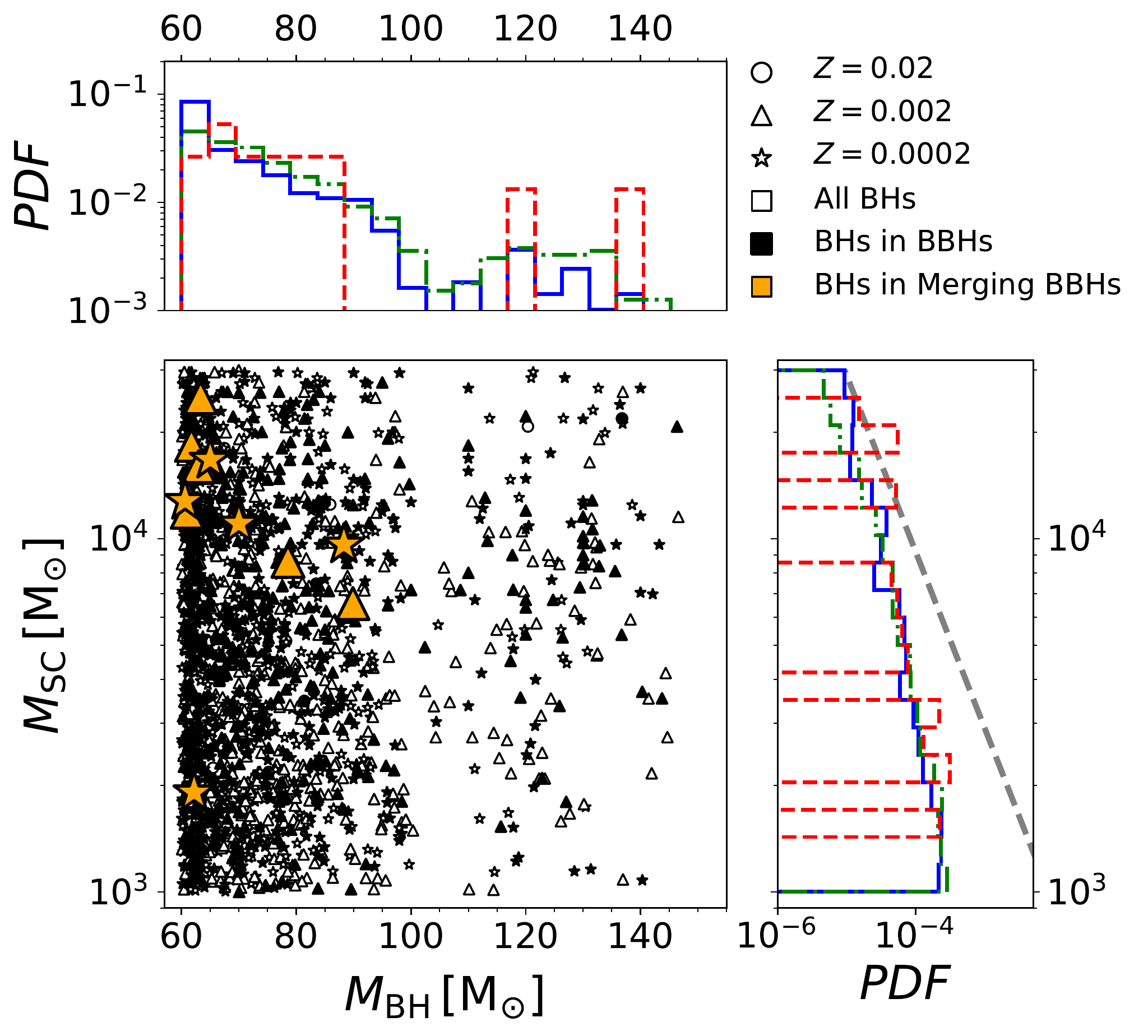,width=8.0cm}
    \caption{Mass of the host star cluster ($M_{\rm SC}$) versus the mass $M_{\rm BH}$ of a BH in the PI gap. Marginal histograms show the distribution of $M_{\rm SC}$ (on the $y-$axis) and $M_{\rm BH}$ (on the $x-$axis). Orange and black filled symbols refer to BBHs merging within a Hubble time and to all BBHs, respectively. Open symbols show single BHs.
     Solid blue, dot-dashed green and dashed red histograms refer to $Z=0.0002,$ 0.002 and 0.02, respectively. The grey dashed line shows the mass function of $M_{\rm SC}$ in our simulation set ($dN/dM_{\rm SC}\propto{}M_{\rm SC}^{-2}$).\label{fig:msc}} 
}
\end{figure}

\subsection{Mass distribution}
Figure~\ref{fig:qmtot} shows the mass ratio $q=M_2/M_1$ (where $M_1>M_2$) and the total mass $M_{\rm tot}=M_1+M_2$ of all BBHs that have at least one member in the PI gap. We form BHs with masses in the entire range of the PI gap between $\sim{}60-150$ M$_\odot$, with a preference for masses around $60-70$ M$_\odot$.

Values of mass ratio $q\gtrsim{}0.4$ are the most likely, but we find binaries with $q$ as low as $\sim{}0.04$. 
The binary with the smallest value of $q$ has secondary mass $M_2\sim{}4.2$ M$_\odot$. The largest secondary mass is $M_2\sim{}110$ M$_\odot$. 
Overall, binaries hosting a BH with mass in the gap have lower mass ratios than other BBHs (see Figure~7 of \citealt{dicarlo2019}, where we show that the vast majority of BBHs in young SCs have $q\sim{}0.9-1$). 

Figure~\ref{fig:msc} shows the mass of the host SC as a function of the mass of BHs in the PI gap (here we include also BHs that remain single). BHs in the mass gap form more efficiently in massive young SCs, where dynamics is more important. 
Ten out of eleven merging BBHs are hosted in star clusters with $M_{\rm SC}>6000$ M$_\odot$, among the most massive young SCs in our sample. 

\begin{table}
\begin{center}
\caption{Fraction of BHs, BBHs and merging BBHs with mass in the PI gap.  \label{tab:table2}} \leavevmode
\begin{tabular}[!h]{ccccc}
\hline
$Z$ &  $f_{\rm PI,\,{}BHs}$ & $f_{\rm PI,\,{}BBHs}$  & $f_{\rm PI,\,{}GW}$  & $p_{\text{det}}^{\text{PI}}$\\
\hline
0.0002 & $5.6 \pm 0.3$~\% & $20.8 \pm 1.7$~\% & $2.2 \pm 1.9$~\% & 11.2~\%\\
0.002 & $1.5 \pm 0.1$~\% & $9.6 \pm 1.0$~\% & $2.1 \pm 1.6~\%$ & 10.0~\% \\
0.02 & $0.1 \pm 0.04$~\% & $0.5 \pm 0.5$~\% & 0.0~\% & 0.0~\%\\
S2020 & -- & -- & 0.5~\% & 5.3~\% \\
\hline
\end{tabular}
\end{center}
\begin{flushleft}
  \footnotesize{Column~1 ($Z$): progenitor's metallicity; S2020 indicates that we accounted for progenitor's metallicity evolution as a function of redshift, as described in \cite{santoliquido2020}}; column~2 ($f_{\rm PI,\,{}BHs}$): percentage of BHs with mass in the PI gap with respect to all simulated BHs at a given $Z$; column~3 ($f_{\rm PI,\,{}BBHs}$): percentage of BBHs that have at least one member with mass in the PI gap with respect to all BBHs  at a given $Z$ formed by the end of the simulations. column~4 ($f_{\rm PI,\,{}GW}$): percentage of merging BBHs that have at least one member with mass in the PI gap with respect to all merging BBHs  at a given $Z$ (a merging BBH is defined as a BBH which merges in less than a Hubble time by GW emission). Errors on $f_{\rm PI,\,{}BHs}$, $f_{\rm PI,\,{}BBHs}$ and $f_{\rm PI,\,{}GW}$ correspond to 95\% credible intervals on binomial distributions, using a Wald method for approximation. Column~5 ($p_{\text{det}}^{\text{PI}}$): percentage of detectable BBH mergers  that have at least one member with mass in the PI gap with respect to all detectable BBH mergers  at a given $Z$ (see equation~\ref{eq:pdet}).
\end{flushleft}
\end{table}

\subsection{Merger and detection efficiency}
We find that only $\sim{}0-2.2$~\% of all merging BBHs have at least one member with mass in the PI gap, depending on metallicity. However, these systems are  more massive than other merging BBHs, thus they have a higher detection chance. To properly take into account these selection effects, we followed a similar approach as in \cite{finn1992}, \cite{dominik2015} and \cite{bouffanais2019}. 

We associate to each mock source (in our catalogue of 534 merging BBHs) the optimal signal-to-noise ratio (SNR) $\rho_{opt}$ that corresponds to the case where the source is optimally oriented and located in the sky. Since real-life sources have different orientations and locations, we then reweigh the SNR as $\rho = \omega \times \rho_{\rm opt}$, where $\omega$ takes randomly generated values between 0 and 1, and the probability of detecting a source is given by
\begin{equation}
p_{\text{det}} = 1 -F_{\omega}(\rho_{\text{thr}} / \rho_{\rm opt}).
\end{equation}
In this equation, $F_{\omega}$ is the cumulative function of $\omega$ and $\rho_{\text{thr}}$ is a detection threshold. We use $\rho_{\text{thr}} = 8$, that was shown to be a good approximation for a network of detectors \citep{abadie2010,abbottGW150914Rates}. We used the software PyCBC \citep{canton2014,usman2015} to generate both the waveforms (IMRPhenomB with zero spins) and the noise power spectral densities of advanced LIGO at design sensitivity \citep{abbott2018}, and the package gwdet \citep{gwdet_ref} to evaluate the function $F_{\omega}$.

From there, we ran two different analysis: one where each set of metallicity is treated independently, and the other where we combine them together using a model describing redshift and metallicity evolution. In the first scenario, for each metallicity set we construct a catalogue of $10^6$ sources where the masses are drawn uniformly from the catalogue and redshifts are drawn uniformly in comoving volume between 0 and 1. In the second scenario, we first compute the merger rate at the detector as a function of redshift, by making use of the {cosmo$\mathcal{R}$ate} code \citep{santoliquido2020}. In particular, following \cite{santoliquido2020}, we assume that all stars form in young SCs, we account for the cosmic star formation rate  \citep{madau2017} and for the stellar metallicity evolution \citep{decia2018}, and we take cosmological parameters from \cite{planck2016}. From there, we build a catalogue of $10^{6}$ sources, by making use once again of the cosmo$\mathcal{R}$ate code \citep{santoliquido2020},  to have the distribution of masses as a function of redshift.

Finally,  to obtain the probability of detecting a source with at least one component in the PI mass gap, we computed  the following quantity for both analyses:
\begin{equation}\label{eq:pdet}
p_{\text{det}}^{\text{PI}} = \sum_{i \in \text{PI}} p^i_{\text{det}} / \sum_{j} p^j_{\text{det}},
\end{equation}
where the sum in the numerator is done only over sources where at least one component lies in the mass gap while the sum in the denominator is done over all sources in our catalogue of merging BBHs. 

We find  $p_{\text{det}}^{\text{PI}} = 0-11$\%, depending on metallicity (see the last column of Table~\ref{tab:table2}). This means that, under our assumption that all stars form in young SCs, up to $11$\% of all BBHs detected by LIGO-Virgo at design sensitivity have at least one component in the PI mass gap. If we assume a model-dependent BBH merger rate evolution with redshift (based on the cosmic star formation rate density and on the average metallicity evolution, \citealt{santoliquido2020}), we find $p_{\text{det}}^{\text{PI}} \sim{}5$~\%, under the assumption that all cosmic star formation takes place in young SCs like the ones we simulated.

\section{Conclusions}
Pair instability (PI) and pulsational PI prevent the formation of BHs with mass between $\sim{}60$ and $\sim{}150$ M$_\odot$ from single stellar evolution. However, binary evolution processes (such as stellar mergers) and dynamical processes might allow the formation of BHs with masses in the gap.

Here, we investigate the possibility that BHs with mass in the gap form  through stellar mergers and multiple stellar mergers in young SCs. The merger between an evolved star (a giant with a well developed helium core) and a main sequence star can give birth to a BH with mass in the gap, provided that the star collapses before its helium core grows above $\sim{}32$ M$_\odot$. In our simulations, these stellar mergers are facilitated by the SC environment: dynamical encounters perturb a binary star, affecting its orbital properties and increasing the probability of a merger between its components. Some massive stars even undergo runaway collisions: they go through multiple mergers over few Myrs. When a BH with mass in the PI gap forms in this way, it is initially a single object. If it remains in the SC, it can acquire a new companion through dynamical exchanges. In contrast, BHs that form via stellar mergers in the field remain single BHs. Moreover,  BHs with masses $>60$ M$_\odot$ are much harder to form in isolated binaries, because non-conservative mass transfer peels-off the primary before the merger. Dynamical encounters perturb the binary and induce a fast merger without episodes of mass transfer.

We have investigated the formation and the dynamical evolution of BHs with masses in the gap through $10^4$ direct N-body simulations of young SCs with metallicity $Z=0.0002,$ 0.002 and 0.02 and with total mass between $10^3$ and $3\times{}10^4$ M$_\odot$. Hence, we focused on relatively small young SCs. At the end of our simulations, $\sim{}5.6$~\%, $\sim{}1.5$~\% and $\sim{}0.1$~\% of all BHs have mass in the PI gap for metallicity $Z=0.0002$, 0.002 and 0.02, respectively. Metal-poor stars are more efficient in producing these BHs, because they lose less mass by stellar winds. In our simulations, we do not include prescriptions for BH spins, because the connection between the spin of the progenitor star and the spin of the BH is highly uncertain (see e.g. \citealt{heger2005,lovegrove2013,belczynski2017spin,qin2018,qin2019,fuller2019,fuller2019b}). We can speculate that stellar mergers spin up the progenitor stars, but we cannot tell whether this spin-up translates into a higher BH spin.

The treatment of the merger of two stars in our simulations is simplified: we assume no mass loss and no chemical mixing during the merger and we require that the merger product reaches hydro-static equilibrium instantaneously. The merger product is rejuvenated according to \cite{hurley2002} simple prescriptions. 
Hydro-dynamical simulations of a stellar merger are required in order to have a better understanding of the final outcome. Thus, our results should be regarded as an upper limit to the formation of BHs in the PI mass gap via stellar mergers.

In our simulations, several BHs with masses in the gap end up forming a BBH through dynamical exchanges. BBHs having at least one component in the mass gap are $\sim{}20.6$~\%, $\sim{}9.8$~\% and $\sim{}0.5$~\% of all BBHs in our simulations, for metallicity $Z=0.0002,$ 0.002 and 0.02, respectively. Thus, BHs with masses in the gap are quite efficient in forming BBHs. The total masses of these BBHs are typically around $M_{\rm TOT}\sim{}90-130$ M$_\odot$ and the most likely mass ratios are $q\gtrsim{}0.4$.

In our simulations, $\sim{}2.1$~\% ($\sim{}2.2$~\%) of all BBHs merging within a Hubble time have at least one component in the mass gap for metallicity $Z=0.002$ ($Z=0.0002$). We find no merging BBHs in the mass gap at solar metallicity. Merging BBHs in the mass gap form preferentially in the most massive SCs we simulate ($M_{\rm SC}\ge{}6000$ M$_\odot$) Hence, BBH mergers in the mass gap might be even more common in higher mass SCs (e.g. globular clusters) than the ones we simulate. Since merging BBHs in the mass gap form through dynamical exchanges, their spins will be isotropically oriented with respect to the orbital angular momentum of the binary system.

Finally, we calculate the probability that advanced LIGO and Virgo at design sensitivity detect the merger of BBHs in the mass gap. Modelling the dependence of the merger rate on the cosmic star formation rate density and metallicity evolution \citep{santoliquido2020}, we predict that $\sim{}5$~\% of all BBH mergers detected by LIGO and Virgo at design sensitivity have at least one component in the PI mass gap, under the assumption that all stars form in young SCs.  
If the proposed mechanism to form BHs in the mass gap is actually at work, the LIGO-Virgo collaboration might be able to witness these events in the next few years.

\section*{Acknowledgments}
We thank the anonymous referee for their useful comments. We thank Mark Gieles, the internal P\&{}P reviewer of the LVC. UNDC acknowledges financial support from Universit\`a degli Studi dell'Insubria through a Cycle 33rd PhD grant.
MM, YB, NG and FS   acknowledge financial support by the European Research Council for the ERC Consolidator grant DEMOBLACK, under contract no. 770017.
MS acknowledges funding from the European Union's Horizon 2020 research and innovation programme under the Marie-Sklodowska-Curie grant agreement No. 794393.
AB  acknowledges support by PRIN MIUR 2017 prot.20173ML3WW 002 "Opening the ALMA window on the cosmic evolution of gas, stars and supermassive black holes".
This work benefited from support by the International Space Science Institute (ISSI), Bern, Switzerland,  through its International Team programme ref. no. 393  {\it The Evolution of Rich Stellar Populations \& BH Binaries} (2017-18).
 

\section*{Data Availability}
The data underlying this article will be shared on reasonable request to the corresponding authors.
\bibliography{./bibliography}
\end{document}